\documentstyle[12pt]{article}

\newcommand{\EQ}{\begin{equation}}
\newcommand{\EN}{\end{equation}}
\newcommand{\bea}{\begin{eqnarray}}
\newcommand{\ena}{\end{eqnarray}}
\newcommand{\vs}[1]{\vspace{#1 mm}}

\renewcommand{\d}{\delta}
\newcommand{\e}{\epsilon}
\renewcommand{\t}{\theta}

\newcommand{\shalf}{\frac{1}{2}}
\newcommand{\pa}{\partial}

\newcommand{\nn}{\nonumber \\}

\begin{document}

\topmargin 0pt
\oddsidemargin 5mm

\renewcommand{\Im}{{\rm Im}\,}
\newcommand{\NP}[1]{Nucl.\ Phys.\ {\bf #1}}
\newcommand{\PL}[1]{Phys.\ Lett.\ {\bf #1}}
\newcommand{\CMP}[1]{Comm.\ Math.\ Phys.\ {\bf #1}}
\newcommand{\PR}[1]{Phys.\ Rev.\ {\bf #1}}
\newcommand{\PRL}[1]{Phys.\ Rev.\ Lett.\ {\bf #1}}
\newcommand{\PTP}[1]{Prog.\ Theor.\ Phys.\ {\bf #1}}
\newcommand{\PTPS}[1]{Prog.\ Theor.\ Phys.\ Suppl.\ {\bf #1}}
\newcommand{\MPL}[1]{Mod.\ Phys.\ Lett.\ {\bf #1}}
\newcommand{\IJMP}[1]{Int.\ Jour.\ Mod.\ Phys.\ {\bf #1}}

\begin{titlepage}
\setcounter{page}{0}
\begin{flushright}
NBI-HE-94-20\\
March 1994\\
hep-th/9403150
\end{flushright}

\vs{8}
\begin{center}
{\Large A HIERARCHY OF SUPERSTRINGS}

\vs{15}
{\large Fiorenzo Bastianelli\footnote{e-mail address:
 fiorenzo@nbivax.nbi.dk}}\\
{\em The Niels Bohr Institute, Blegdamsvej 17, DK-2100 Copenhagen \O,
Denmark}

\vs{8}
{\large Nobuyoshi Ohta\footnote{e-mail address:
ohta@nbivax.nbi.dk, ohta@fuji.wani.osaka-u.ac.jp}$^,$\footnote{
Permanent address and address after April 1, 1994: Department of Physics,
Osaka University, Machikaneyama-cho 1-16, Toyonaka, Osaka 560, Japan}}\\
{\em NORDITA, Blegdamsvej 17, DK-2100 Copenhagen \O, Denmark}\\

\vs{8}
{\large Jens Lyng Petersen\footnote{e-mail address:
 jenslyng@nbivax.nbi.dk}}\\
{\em The Niels Bohr Institute, Blegdamsvej 17, DK-2100 Copenhagen \O,
Denmark}
\end{center}

\vs{8}
\centerline{{\bf{Abstract}}}

We construct a hierarchy of supersymmetric string theories by
showing that the general $N$-extended superstrings may be viewed
as a special class of the $(N+1)$-extended superstrings.
As a side result, we find a twisted $(N+2)$ superconformal
algebra realized in the $N$-extended string.

\end{titlepage}
\newpage
\renewcommand{\thefootnote}{\arabic{footnote}}
\setcounter{footnote}{0}

Recently a remarkable discovery has been made that string theories
may be interpreted as kinds of spontaneously broken phases of those with
higher world-sheet symmetries~\cite{BV}. In particular, it has been shown
that the $N=0~(N=1)$ strings can be viewed as a special class
of vacua for the $N=1~(N=2)$ superstrings~[1-5].
It has also been shown that $N=2$ superstrings can be regarded as different
phases of $N=4$ strings~\cite{BOP}. In this process it has been observed
that a similar structure exists beyond $N=2$~\cite{BOP,BO}, and it has
been speculated that it is possible to embed the general $N$ superstrings into
the $(N+1)$ superstrings for $N\geq 2$~\cite{BERK,BV,FO}.

The purpose of this paper is to realize explicitly the $N$-extended
superstrings as special choices of the vacua in the $(N+1)$-extended
superstrings, and to give a simple and explicit proof that this $(N+1)$
formulation is equivalent to the $N$ superstring.
By $N$-extended superstrings we mean those based on the linear algebras
found in ref.~\cite{N4S}.
Our results apply to general $N$ superstrings for $N\geq 2$, and hence imply
that there is an infinite hierarchy of superstrings.
We also find a twisted $(N+2)$ superconformal symmetry realized in the
$N$ string.

We will use the $N$-extended superspace to describe these theories.
Our conventions follow closely those of ref.~\cite{SCH}:
$Z=(z,\t^i)$ with $i=1,2,\dots,N$ denotes the $N$-extended
super-coordinates and
\bea
& & D_i \equiv \pa_{\t^i} + \t^i\pa_z \ , \qquad
\{ D_i, D_j \} = 2 \d_{ij} \pa_z \ , \nn
& & \t^i_{12} \equiv \t^i_1 - \t^i_2 \ , \qquad
z_{12} \equiv z_1 - z_2 - \t^i_1 \t^i_2 \ , \nn
& & \t^N \equiv \frac{1}{N!}
\e^{j_1\dots j_N} \t^{j_1} \cdots \t^{j_N} \ , \qquad
\t^{N-i} \equiv \frac{1}{(N-1)!}
\e^{j_1 \cdots j_{N-1} i} \t^{j_1} \cdots \t^{j_{N-1}}\ .
\ena

The $N$ superconformal algebra (SCA) is generated by the super stress
tensor $T$, which satisfies the operator product expansion (OPE)
\EQ
T(Z_1) T(Z_2) \sim  \frac{\t^N}{z^2} \left( 2 - \frac{N}{2} \right) T(Z_2)
 + \frac{\t^{N-i}}{z} \frac{1}{2} D_i T(Z_2) +
\frac{\t^N}{z} \pa T(Z_2) \ ,
\label{scan}
\EN
where we have used the shorthand notation $z \equiv z_{12}$ and
$\t^N \equiv \t^N_{12}$. Central extensions can appear
for $N \leq 4$~\cite{SCH}. However, we will mainly focus on the cases
which do not have central extensions ($N\geq 5$), or have a vanishing value
for the critical central charges ($N=3,4$).
We will comment on the case $N=2$ later on.
One can construct a string theory by using the $N$ SCA as the
(chiral) constraint algebra defining the states of the string.
This information is encoded in a nilpotent
BRST operator, constructed as the superspace integral of the BRST current
$Q_N$,
\EQ
Q_{N} = C_t \left (T + \shalf T(B_t,C_t) \right)  ,
\label{1}
\EN
where the reparametrization ghosts $(B_t, C_t)$ are fields of
spin $(2 -\frac{N}{2},-1)$, $C_t$ being anticommuting for any $N$,
and $T(B,C)$ denotes the super stress tensor of an arbitrary
$BC$ system
\EQ
T(B,C) = (-1)^{N(\e_C+1)} \left ( - \lambda_B B \pa C +
\lambda_C \pa B C + \frac{(-1)^{\e_B}}{2} D_i B D_i C \right ).
\EN
In this last formula, $\lambda$ and $ \e$ denote
the conformal spin and Grassmann character of a field
(even $\e$ for bosonic fields and odd $\e$ for fermionic ones),
respectively, and satisfy
\EQ
\lambda_B + \lambda_C = 1 - \frac{N}{2} \ , \qquad \e_B + \e_C = N \ .
\label{ss}
\EN
We also use the following correlator for a BC system
\EQ
 C(Z_1)B(Z_2)\sim  \frac{\t_{12}^N}{z_{12}} \ .
\label{cor}
\EN
The OPE of the BRST currents with itself can be computed and is given by
\EQ
Q_N(Z_1) Q_N(Z_2) \sim  - \frac{(-1)^{N}}{4}\frac{\t^N}{z}
D_i( (D_iC_t) Q_N) (Z_2)\ .
\EN
However, one can improve the BRST current by a total derivative term
\EQ
\tilde Q_N = Q_N - \frac{1}{4} D_i ( (D_i C_t) B_t C_t) \ ,
\label{2}
\EN
to make it completely nilpotent
\EQ
\tilde Q_N(Z_1) \tilde Q_N(Z_2) \sim  0 \ .
\EN
This result will be helpful in constructing the general embedding to be
discussed shortly and, more generally, to uncover a twisted $(N+2)$
SCA realized in the $N$ string.

The $(N+1)$ SCA in $N$ superfields is given by
\bea
T(Z_1) T(Z_2) &\sim&  \frac{\t^N}{z^2} \left( 2 - \frac{N}{2} \right) T(Z_2)
 + \frac{\t^{N-i}}{z} \frac{1}{2} D_i T(Z_2) + \frac{\t^N}{z} \pa T(Z_2)
\ ,\nn
T(Z_1) G(Z_2) &\sim& \frac{\t^N}{z^2}\left(\frac{3}{2} - \frac{N}{2} \right)
  G(Z_2) + \frac{\t^{N-i}}{z} \frac{1}{2} D_i G(Z_2)
 + \frac{\t^N}{z} \pa G(Z_2) \ ,\nn
G(Z_1) G(Z_2) &\sim& \frac{\t^N}{z} 2 T(Z_2) \ .
\label{sca}
\ena
This algebra can be written as in (\ref{scan}) using
$(N+1)$ superfields and with the $(N+1)$ stress tensor defined in an
obvious notation by $T_{N+1} \equiv G_N/2 +\t_{N+1}T_N$.
Again, all the information on physical states of the $(N+1)$ string theory
is encoded in the BRST current for the $(N+1)$ algebra
\EQ
Q =  C_t T + C_g G  + C_t \left( \shalf T(B_t,C_t) + T(B_g,C_g)
\right) - C_g^2 B_t \ ,
\label{brst}
\EN
where the anticommuting reparametrization ghosts $C_t$
and the commuting supersymmetry ghost $C_g$ have
spins $ -1 $ and $-\shalf$, respectively.

The embedding of the $N$ string into the $(N+1)$ string is achieved as
follows. We first take an arbitrary matter background $T_m$ for the $N$
string which satisfies eq.~(\ref{scan}). We then add to this a
$BC$ system denoted by $(\eta,\xi)$, with $\xi$ anticommuting
and with spin $-\shalf$. The spin and statistics of $\eta$ follow
then from eq. (\ref{ss}).
With these ingredients at hand, we can construct the following background for
the $(N+1)$ string:
\bea
T &=& T_m  + T( \eta, \xi) \  ,\nn
G &=& \eta + \xi T_m + \left(\frac{N}{2} - 1 \right ) \xi
\eta \pa \xi  + \frac{(-1)^N}{4} \eta (D_i\xi)^2
 - \frac{1}{2} (D_i \eta)(D_i \xi) \xi \ ,
\label{rea}
\ena
where $T(\eta,\xi)$ is the super stress tensor for the $(\eta,\xi)$
system. One can show that these generators indeed satisfy the operator
products given in eq.~(\ref{sca}).

This construction is easy to understand. The supersymmetry generator $G$
has the structure $G\sim \eta + \tilde Q_N(\eta,\xi)$,
where $\tilde Q_N(\eta,\xi)$ is the improved BRST current of eq.~(\ref{2})
with the reparametrization ghosts $(B_t,C_t)$ replaced by $(\eta,\xi)$.
Since the improved BRST current is fully nilpotent, only the
contractions between $\eta$ and $\tilde Q_N(\eta,\xi)$ are
nonvanishing and generate the super stress tensor $T$, as can be
imagined recalling the usual BRST algebra.
Note, however, that the spin of the $(\eta,\xi)$ system gets
modified to the values $(\frac{3}{2} -\frac{N}{2}, -\shalf)$
while $(B_t,C_t)$ had originally spin $(2 - \frac{N}{2},-1)$.
Actually, this construction can be extended to reveal a twisted
$(N+2)$ algebra in the $N$ string.
We will postpone the description of such an algebra to the end of the paper.
Note also that one can think of
$(\eta,\xi)$ as fields with the same quantum number of $(B_g,C_g)$
but opposite statistics, so that they will cancel each other through the
BRST quartet mechanism.

The BRST current corresponding to the particular background just
constructed is obtained by substituting eq.~(\ref{rea}) into eq.~(\ref{brst})
\bea
Q_{N+1} &=&  C_t \left(T_m + T(\eta,\xi)\right)\nn
&+& C_g \left( \eta +\xi T_m + \left(\frac{N}{2} - 1 \right ) \xi
\eta \pa \xi  + \frac{(-1)^N}{4} \eta (D_i \xi)^2
 - \frac{1}{2}(D_i \eta)(D_i \xi) \xi \right)\nn
&+&  C_t \left( \shalf T(B_t,C_t) + T(B_g,C_g) \right) - C_g^2 B_t \ .
\label{npobrs}
\ena
To prove that this particular class of $(N+1)$ string theories is
equivalent to the $N$ string, we perform a canonical
transformation to map $Q_{N+1}$ onto $Q_N + Q_{top}$, where
$Q_N$ was given in (\ref{1}) and
\EQ
Q_{top}= C_g \eta
\label{top}
\EN
gives the BRST charge for a trivial topological sector.
This was the strategy already employed
in refs.~\cite{IK,FB,OP,BOP}, where a canonical transformation was used to
map the $(N+1)$ supersymmetric formulation of the $N$ string onto the
standard formulation for $N=0,1,2$.
In particular, we follow ref.~\cite{FB} which presented
the canonical transformation factorized in three parts,
each of which is of simpler construction and interpretation.
The first canonical transformation we perform is generated by
\EQ
R_1 = \oint
B_t C_g \xi \ ,
\label{ftra}
\EN
where $\oint $ denotes superspace integration.
The property of $R_1$ is to make $T_m$ inert under the extra
supersymmetry generated by $G$.
We find that this transformation acts as follows:
\bea
Q_{N+1}' \equiv e^{R_1}Q_{N+1}e^{-R_1}
 &=&  C_t \left(T_m + \shalf T(B_t,C_t) + T(B_g,C_g)
 + T(\eta,\xi)  \right) \nn
&-& C_g \xi T(B_g,C_g) + C_g \left( \eta + \shalf \xi\eta\pa\xi
 + \frac{(-1)^N}{4} \eta (D_i \xi)^2 \right) .
\label{fres}
\ena
Here we consistently drop total derivative terms which may appear
on the right hand side of this equation, since they will not affect
the BRST charge. Next, we perform a second transformation generated by
\EQ
R_2 = \oint \left[
\left(\frac{N}{4}-\shalf\right)B_g C_g \xi \pa \xi
 + \frac{1}{4} C_g \xi D_i \xi D_i B_g  \nn
 + \frac{1}{8} ( \xi \eta  - C_g B_g) (D_i \xi)^2 \right]
\label{stra}
\EN
to simplify the BRST algebra in the $(B_g,C_g,\eta,\xi)$ topological sector.
This casts the BRST charge into
\EQ
Q_{N+1}'' \equiv e^{R_2}Q_{N+1}'e^{-R_2}
 = C_t \left(T_m + \shalf T(B_t,C_t) + T(B_g,C_g)
 + T(\eta,\xi) \right) + C_g \eta .
\label{sres}
\EN
A final transformation generated by
\EQ
R_3 = \oint \left[
(-1)^N C_t \left ( \left (\frac{3}{2} - \frac{N}{2} \right )
B_g \pa \xi + \shalf \pa B_g \xi
- \frac{(-1)^N}{2} D_i B_g D_i \xi \right )\right] \  ,
\label{ttra}
\EN
is used to decouple the BRST reparametrizations from the topological
sector. It maps modulo total derivatives (\ref{sres}) onto $Q_N + Q_{top}$,
as given in eqs.~(\ref{1}) and (\ref{top}).

Obviously the BRST charges constructed out of
$Q_N$ and $Q_{top}$ commute with each other
and are nilpotent, giving separate conditions on the physical states.
The topological charge $\oint Q_{top}$ imposes the condition that the fields
$(B_g,C_g,\eta,\xi)$ fall into a quartet representation of the charge and
decouple from the physical subspace, and we are left with the degrees
of freedom of the $N$ superstring only. Thus the $(N+1)$ string propagating in
the background  described by eqs.~(\ref{rea}) is equivalent to the $N$ string.
Note also that our similarity transformations manifestly preserve the
operator algebra.

For the embedding $N=2$ into $N=3$, one has to recall that there is a central
extension $\sim c/(3 z^2)$ appearing on the right hand side of
eq.~(\ref{scan}) with the critical value $c=6$. However, this central
charge is balanced by a contribution $c=-6$ due to the $(\eta,\xi)$
system. We have
checked that our realization (\ref{rea}) as well as our canonical
transformations work also in this case. Thus our construction
is valid for arbitrary $N \geq 2$ and, when combined with
the embeddings found in ref.~\cite{BV}, leads to the amazing result that
there exists an infinite hierarchy of superstrings.

We have also found that out of $T_m$ and $(\eta,\xi)$ one can construct
an $(N+2)$ SCA:
\bea
T(Z_1) T(Z_2) &\sim&  \frac{\t^N}{z^2}\left( 2-\frac{N}{2} \right)T(Z_2)+
\frac{\t^{N-i}}{z} \frac{1}{2} D_i T(Z_2) +
\frac{\t^N}{z} \pa T(Z_2) \ ,\nn
T(Z_1) G^\pm(Z_2) &\sim&  \frac{\t^N}{z^2}
\left( \frac{3}{2} - \frac{N}{2} \right) G^\pm(Z_2)
 + \frac{\t^{N-i}}{z} \frac{1}{2} D_i G^\pm(Z_2) +
\frac{\t^N}{z} \pa G^\pm(Z_2) \ ,\nn
T(Z_1) J(Z_2) &\sim&  \frac{\t^N}{z^2}
\left(1 - \frac{N}{2} \right) J(Z_2) +
\frac{\t^{N-i}}{z} \frac{1}{2} D_i J(Z_2) +\frac{\t^N}{z}\pa J(Z_2) \ ,\nn
G^+(Z_1) G^-(Z_2) &\sim& \frac{\t^N}{z^2}\left(1-\frac{N}{2} \right)J(Z_2) +
\frac{\t^{N-i}}{z} \frac{1}{2} D_i J(Z_2) +  \frac{\t^N}{z}
\left(T + \shalf \pa J \right)(Z_2) \ , \nn
J(Z_1) G^\pm(Z_2) &\sim& \pm \frac{\t^N}{z}G^\pm(Z_2) \ .
\ena
The realization is given by
\bea
T &=& T_m  + T( \eta, \xi) \ ,\nn
G^+ &=& \eta \ ,\nn
G^- &=&  \xi T_m + \left(\frac{N}{2} - 1 \right ) \xi
\eta \pa \xi  + \frac{(-1)^N}{4} \eta(D_i \xi)^2
 - \frac{1}{2}(D_i \eta)(D_i \xi) \xi \ ,\nn
J &=& \eta\xi \ .
\ena
This is the twisted $(N+2)$ SCA present in the $N$ string. In fact,
the fields $(\eta,\xi)$ correspond to the ghosts $(B_t,C_t)$,
but their stress tensor is twisted in order to shift the spins from
the value  $(2 - \frac{N}{2},-1)$ to the value
$(\frac{3}{2} -\frac{N}{2}, -\shalf)$.
For $N=2$, this reduces to the special case known in the
literature~\cite{GR}. That such a construction was possible was
suggested by Berkovits as reported in the first of refs.~\cite{GR}.
One can use it to construct a consistent background for the
$(N+2)$ string, and by performing canonical transformations
it is possible to show that this formulation is also equivalent
to the $N$ string,
as explicitly demonstrated in ref.~\cite{BOP} for the case $N=2$.

Concerning linear superalgebras,
it remains to be seen if one can formulate the small $N=4$
superstring~\cite{N4S} with $SU(2)$ symmetry as
a particular class of the large $N=4$ superstring and
the general large $N=4$ superstrings depending on a free parameter
$x$~\cite{SCH,STV} as special backgrounds for the $N=5$ strings.
Another line of investigation is to try to embed strings into $W$-strings.
For recent advances in this latter topic, see refs.~\cite{W}.
We hope to discuss these issues elsewhere.

\vs{10}
\noindent
{\it Acknowledgements}

We would like to thank Nathan Berkovits for valuable suggestions.
N. O. would also like to thank Paolo Di Vecchia for discussions,
support and kind hospitality at NORDITA, where this work was done.

\end{document}